# Substrate patterning using regular macroporous block copolymer monoliths as sacrificial templates and as capillary microstamps

*Leiming Guo\* Michael Philippi, Martin Steinhart\**

Dr. L. Guo, M. Philippi, Prof. M. Steinhart
Institut für Chemie neuer Materialien, Universität Osnabrück, Barbarastr. 7, 49069 Osnabrück,
E-mail: leiming.guo@uni-osnabrueck.de; martin.steinhart@uni-osnabrueck.de

**Abstract**

Polystyrene-*block*-poly(2-vinylpyridine) (PS-*b*-P2VP) monoliths containing regular arrays of macropores (diameter ~1.1 µm, depth ~0.7 µm) at their surfaces are used to pattern substrates by patterning modes going beyond the functionality of classical solid elastomer stamps. In a first exemplary application, the macroporous PS-*b*-P2VP monoliths are employed as sacrificial templates for the deposition NaCl nanocrystals and topographically patterned iridium films. One NaCl nanocrystal per macropore is formed by evaporation of NaCl solutions filling the macropores followed by iridium coating. Thermal PS-*b*-P2VP decomposition yields topographically patterned iridium films consisting of ordered arrays of hexagonal cells, each of which contains one NaCl nanocrystal. For the second exemplary application, spongy-continuous mesopore systems are generated in the macroporous PS-*b*-P2VP monoliths by selective-swelling induced pore generation. Infiltrating the spongy-continuous mesopore systems with ink allows capillary microstamping of continuous ink films with holes at the positions of the macropores onto glass slides compatible with advanced light microscopy. Capillary microstamping can be performed multiple times under ambient conditions without re-inking and without quality deterioration of the stamped patterns. The macroporous PS-*b*-P2VP monoliths are prepared by double replication of primary macroporous silicon molds via secondary polydimethylsiloxane molds.

**Introduction**

Polymer monoliths containing macropores with diameters in the range from several 100 nm to a few μm are a versatile material platform for a diverse variety of applications including separation, catalysis, energy storage, energy conversion, drug delivery and tissue engineering. They may be employed as surfaces exhibiting specific wetting and adhesion properties or as microreactor arrays. Polymer monoliths with surfaces topographically patterned with macropore arrays are accessible by soft molding against polydimethylsiloxane (PDMS) molds[1] or by non-lithographic synthetic routes such as formation of breath figures,[2] spreading polymer solutions on non-solvents,[3] and macroscale spinodal decomposition.[4] Block copolymers (BCPs) consisting of at least two chemically distinct blocks offer additional synthetic pathways to macroporous polymer monoliths. Examples include electric-field-induced BCP patterning,[5] non-solvent-induced phase separation,[6] and breath figure formation on BCP solutions.[7] Of particular interest is the additional functionalization of macroporous polymer monoliths with continuous-spongy mesopore systems; BCPs already contain a mesoscopic fine structure that may serve as precursor for the mesoporous structure level. The chemically distinct blocks of the BCPs are typically incompatible so that they assemble into ordered nanoscopic domain structures.[8] Mesopores can be generated by selectively degrading one of the BCP components[9] or by selective-swelling induced pore generation.[10] For example, bioinspired anti-adhesive surfaces with hierarchical structure characterized by mesopore systems and macroscale topographic surface patterns were prepared by molding solutions of the BCP polystyrene-*block*-poly(2-vinylpyridine) (PS-*b*-P2VP) against PDMS molds derived from microsphere monolayers combined with selective-swelling induced pore generation.[10d]

Beyond the functionalities of conventional elastomeric stamps, the potential of macroporous polymer monoliths as a means to pattern substrates has by and large remained unexplored. Here, we demonstrate that PS-*b*-P2VP monoliths with surfaces topographically patterned with macroscopic monodomains of macropores allow modes of surface patterning not accessible by classical elastomeric specimens. First, we demonstrate that metal-coated macroporous PS-*b*-P2VP monoliths (such as macroporous monoliths of the corresponding homopolymers) containing one NaCl nanocrystal per macropore can be used as sacrificial template; thermal decomposition of the PS-*b*-P2VP yielded metal films topographically patterned with hexagonal arrays of cells replicating the macroporous structure of the sacrified macroporous PS-*b*-P2VP monoliths. Each cell contained one NaCl nanocrystal initially located in the macropore acting as precursor of the corresponding cell. In this work NaCl is used as a test system. However, the precise positioning of nanocrystals on surfaces demonstrated here may be a generic approach to two-dimensional nanocrystal arrays for electronic, optical, and sensor applications.[11] Furthermore, thin corrugated metal films may exhibit peculiar plasmonic properties potentially enabling peculiar light transmission modes.[12] Moreover, corrugated metal films may show anomalous reflection of light.[13] Finally, the cells of the corrugated metal films may be used as electrochemical reaction compartments. In the second application example, we generated spongy-continuous mesopore systems in macroporous PS-*b*-P2VP monoliths by selective-swelling induced pore generation and used the continuous-spongy mesopore systems as ink reservoirs for capillary microstamping. Transfer of the macroporous surface topography by capillary microstamping onto counterpart substrates yielded ink films containing holes at the position of the macropores; the use of the spongy-continuous mesopore system enabled multiple successive capillary microstamping cycles under ambient conditions without deterioration of the quality of the stamped pattern. For example, hydrophilic-hydrophobic micropatterns consisting of a holey hydrophobic film containing arrays of hydrophilic dots may be generated in this way. Hydrophilic-hydrophobic micropatterns compatible with standard light microscopy, in which water drops located on the hydrophilic dots are immobilized by the surrounding hydrophobic films, may be promising lab-on-chip configurations for parallel chemical synteses and screenings.[14] Macroporous PS-*b*-P2VP monoliths with a lattice constant of 1.5 µm and a macropore diameter of 1.1 µm were accessible by double replication of primary macroporous silicon molds.[15] Negative PDMS replicas of the macroporous silicon were used as secondary molds to form macroporous PS-*b*-P2VP monoliths as positive replicas of the primary macroporous silicon molds.

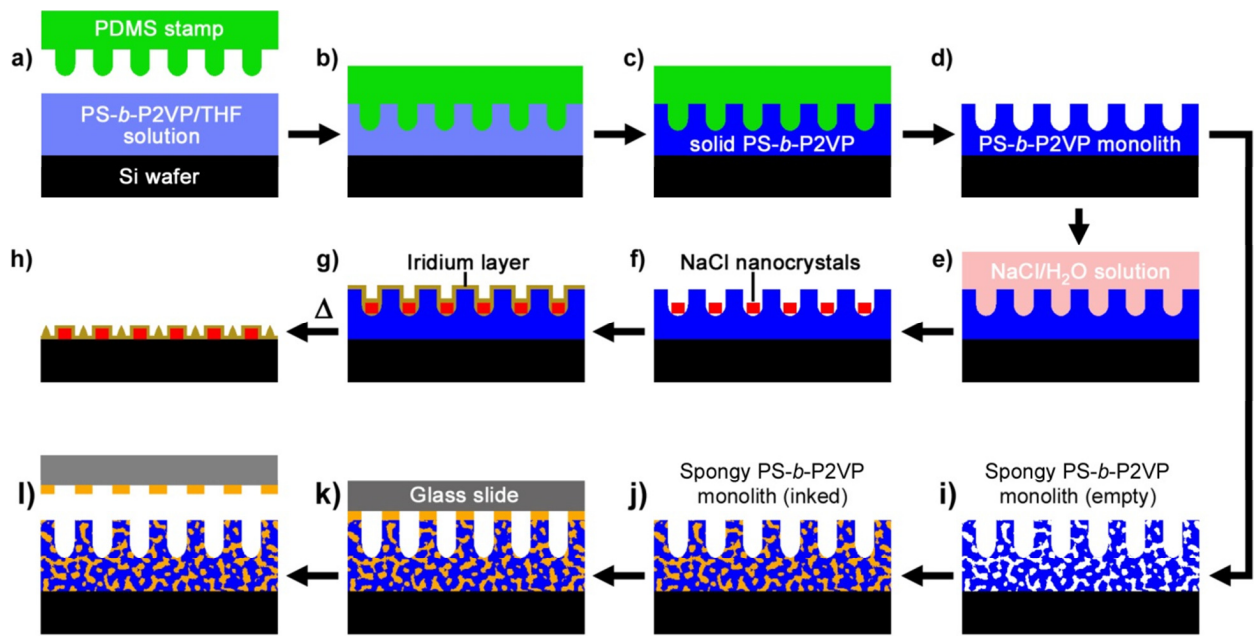

**Figure 1.** Preparation and use of ordered macroporous PS-*b*-P2VP monoliths. a) A PS-*b*-P2VP solution (light blue) is deposited onto a smooth silicon wafer (black). b) The PS-*b*-P2VP solution is imprinted with a secondary PDMS mold (green, negative replica of primary macroporous silicon mold). c) The PS-*b*-P2VP (blue) solidifies after evaporation of the solvent. d) A macroporous PS-*b*-P2VP monolith (blue, positive replica of the primary macroporous silicon mold and negative replica of the secondary PDMS mold) is obtained after detachment of the secondary PDMS mold. e)-h) In a first exemplary application, e) an aqueous NaCl solution (light red) is deposited onto the macroporous PS-*b*-P2VP monolith still attached to the underlying silicon wafer. f) After evaporation of the water, one NaCl nanocrystal (red) forms in each macropore. g) The macroporous PS-*b*-P2VP monolith and the NaCl nanocrystals are coated with a thin iridium layer (ochre). h) Heat treatment results in thermal decomposition of the PS-*b*-P2VP. The remaining iridium skin forms a cellular film on the underlying silicon wafer conserving the hexagonal array geometry of the macroporous PS-*b*-P2VP monolith; in each cell of the iridium skin one NaCl nanocrystal is located. i)-l) In the second application example, i) spongy-continuous mesopore systems are formed in the macroporous PS-*b*-P2VP monoliths by selective-swelling induced pore generation. j) The spongy-continuous mesopore system is filled with ink (orange). The ink-filled spongy macroporous PS-*b*-P2VP monolith is approached to a counterpart substrate, such as a glass slide (light grey). On the areas of the counterpart surface forming contact with the ink-filled spongy macroporous PS-*b*-P2VP monolith ink is deposited. l) After retraction of the spongy macroporous PS-*b*-P2VP monolith, the counterpart surface is patterned with a continuous ink film containing holes at the positions of the macropores. Steps i) –l) can be repeated multiple times under ambient conditions without re-inking.

**Results and discussion**

To prepare macroporous PS-*b*-P2VP monoliths, we used macroporous silicon with a macropore diameter of 1 µm and a macropore depth of 700 nm as primary mold. The macropores in the primary macoporous silicon mold formed extended hexagonal monodomains with a lattice constant of 1.5 µm. To obtain secondary PDMS molds (Supporting Figure S1) as negative replicas of the primary macroporous silicon molds, cross-linkable PDMS prepolymer formulations were molded against the macroporous silicon. The secondary PDMS molds were faithful negative replica of the primary macroporous silicon molds. To prepare the macroporous PS-*b*-P2VP monoliths, we dropped a solution containing 5 wt-% of asymmetric PS-*b*-P2VP (P2VP forms the minority domains) in tetrahydrofuran (THF) onto a silicon wafer (Figure 1a) and placed the secondary PDMS mold on the solution while a pressure of 3.9 kN/m$^2$ was applied (Figure 1b). After evaporation of the THF for at least 15 min (Figure 1c) the secondary PDMS molds were nondestructively detached and macroporous PS-*b*-P2VP monoliths with an area of ~1 cm$^2$ were obtained (Figure 1d). The macropores in the macroporous PS-*b*-P2VP monoliths formed hexagonal monodomains (Figure 2a; see Supporting Figure 2 for a large-field SEM image) replicating the macropore monodomains of the primary macroporous silicon molds. The macropores in the PS-*b*-P2VP monoliths had a diameter ~1.1 µm, whereas the pillar diameter of the secondary PDMS molds amounted to only ~1.0 µm (corresponding to the macropore diameter of the primary macroporous silicon molds). We assume that the secondary PDMS molds were swollen by THF during the molding of the PS-*b*-P2VP/THF solutions. The depth of the macropores in the PS-*b*-P2VP monoliths amounted to ~700 nm and corresponded to the heights of the pillars of the secondary PDMS molds (Supporting Figure S1). A ~1.5 µm thick

PS-*b*-P2VP layer separated the bottoms of the macropores in the PS-*b*-P2VP monoliths from the underlying Si wafer (Figure 2b).

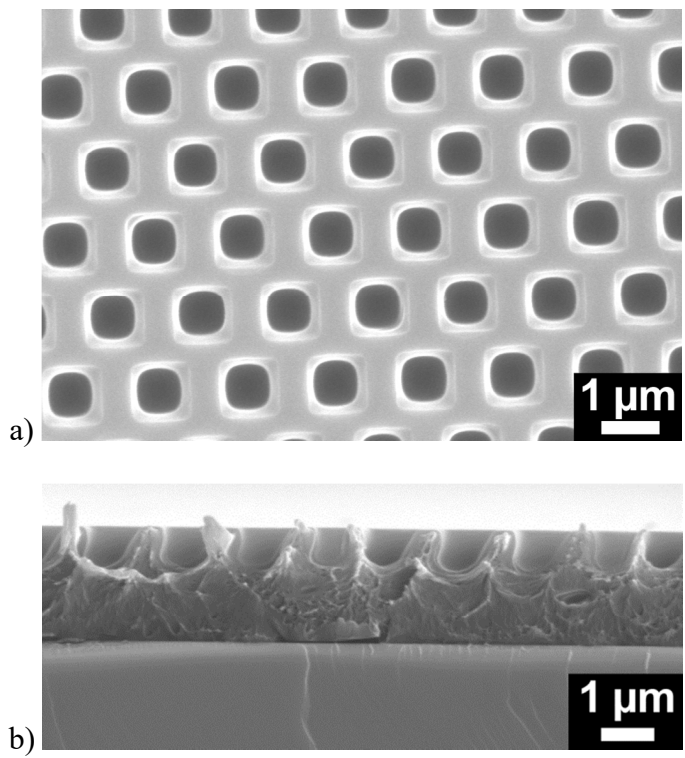

**Figure 2.** Scanning electron microscopy images of macroporous PS-*b*-P2VP monoliths corresponding to Figure 1d obtained with a solution of 5 wt-% PS-*b*-P2VP in THF by imprinting with a secondary PDMS mold (Figure S1) at a pressure of 3.9 kN/m$^2$. a) Top view. b) Cross-section. The macroporous PS-*b*-P2VP monolith on top is connected to a Si wafer at the bottom.

In a first example for substrate patterning with macroporous PS-*b*-P2VP monoliths we used the macroporous PS-*b*-P2VP monoliths as sacrificial templates. It should be noted that in this specific case macroporous monoliths consisting of the corresponding homopolymers could be used as well. At first, we applied the "discontinuous wetting" method[3, 16] to deposit one cubic NaCl nanocrystal with a diameter of ~320 nm into each macropore (Figure 1f and Figure 3a, for a large-field view see Supporting Figure S3). To this end, the macroporous PS-*b*-P2VP monoliths were immersed into an aqueous 2 M NaCl solution for 1 min (Figure 1e) and dried under ambient conditions for 2 h. In the next step we sputter-coated the macroporous PS-*b*-P2VP monoliths containing the NaCl nanocrystals with a ~5 nm thick iridium layer (cf. Figure 1g). Then, the samples were heated to 540°C at a rate of 1K/min and kept at this temperature in air for 3 h. This thermal treatment results in thermal decomposition of the PS-*b*-P2VP.[17] After the heat treatment, the Si wafer was covered with a regularly corrugated iridium film containing a hexagonal pattern of cells separated by iridium rims (for large-field views see Supporting Figure S4a,b) with a heigth of ~160 nm (Supporting Figure S4c). While the cells mark the initial positions of the macropores in the macroporus PS-*b*-P2VP monoliths, the rims are a signature of the PS-*b*-P2VP walls that separated the macropores. Each cell contains one NaCl nanocrystal (Figure 3b,c and Supporting Figure S4b), as schematically displayed in Figure 1h.

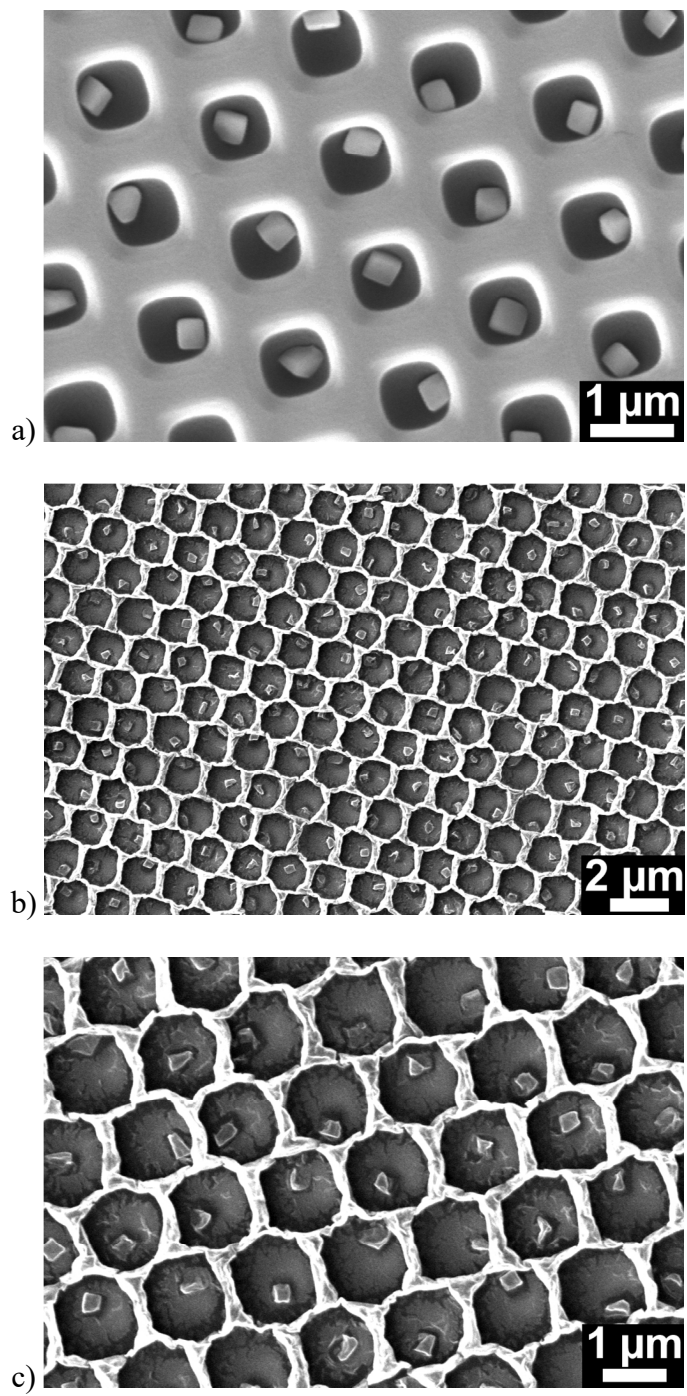

**Figure 3.** Macroporous PS-*b*-P2VP monoliths as sacrificial templates. Scanning electron microscopy image of a) a macroporous PS-*b*-P2VP monolith containing one NaCl nanocrystal in each macropore corresponding to panel f) of Figure 1. b), c) Scanning electron microscopy images of the product obtained by annealing an iridium-coated macroporous PS-*b*-P2VP monolith containing one NaCl nanocrystal in each macropore at 540°C in air. Seen is the remaining iridium skin forming a hexagonal array of cells each of which contains one NaCl nanocrystal corresponding to panel h) of Figure 1.

The second exemplary application of macroporous PS-*b*-P2VP monoliths is the stamping of continuous ink films with holes at the position of the macropores onto counterpart substrates. We have recently reported capillary microstamping of arrays of discrete ink spots using spongy-mesoporous stamps with contact surfaces topographically patterned with rod-like contact elements.[18] While classical stamping procedures such as polymer pen lithography[19] rely on solid stamps requiring adsorption of non-volatile ink components onto the outer stamp surface, capillary microstamping with spongy mesoporous stamps allows either continuous ink supply to the contact surface of the stamps via the spongy mesopore system or the use of the spongy mesopore system as ink reservoir. Multiple capillary microstamping cycles can be carried out manually under ambient conditions without complex alignment procedures, without re-inking and without deterioration of the quality of the stamped pattern. So far, only arrays of discrete dots were accessible by capillary microstamping. It is desirable to expand the range of accessible surface patterns and to generate continuous ink films containing regular arrays of holes by capillary microstamping. For this purpose, we converted solid macroporous PS-*b*-P2VP monoliths into spongy macroporous PS-*b*-P2VP monoliths containing continuous mesopore systems as additional hierarchical structure level by selective-swelling induced pore generation.[10] Selective-swelling induced pore generation can, in principle, be applied to any asymmetric BCP on condition that a solvent selective to the minority blocks exists. In this way, mesoporous PS-*b*-P2VP specimens for membrane separation[20] as well as for metal[10a] and metal oxide deposition[17, 21] were obtained. However, ultrafiltration membranes were also obtained by subjecting polystyrene-*b*-poly(methyl methacrylate) to selective-swelling induced pore generation.[22] Here, the macroporous PS-*b*-P2VP monoliths were treated with ethanol at 70°C for 30 minutes. Osmotic pressure drives ethanol, a solvent selective to P2VP, into the P2VP minority domains. The P2VP blocks adopt stretched conformations to maximize contact to the ethanol molecules. Hence, the P2VP minority domains swell, whereas the glassy PS matrix undergoes morphology reconstruction in order to accommodate the increasing volumes of the swelling P2VP minority domains that eventually form a continuous network. Evaporation of the ethanol results in entropic collapse of the stretched P2VP blocks that relax into their native coiled conformations while the glassy PS matrix fixates the reconstructed morphology. Thus, continuous mesopore systems form in lieu of the swollen P2VP minority domains, the walls of which consist of coiled P2VP blocks[23] (Figure 1i). The mesopore systems generated in the macroporous PS-*b*-P2VP monoliths were open to the environment and had pore sizes ranging from 25 to 60 nm (Figure 4a). They penetrated the entire PS-*b*-P2VP monoliths (Figure 4b). The regular arrays of macropores were retained during selective-swelling induced pore generation (for a large-field SEM image see Supporting Figure S5). The thickness of the spongy macroporous PS-*b*-P2VP monoliths and the depth of the macopores nearly doubled to 3.9 µm and 1.4 µm because of the volume expansion in the vertical direction during selective-swelling induced pore formation.[24]

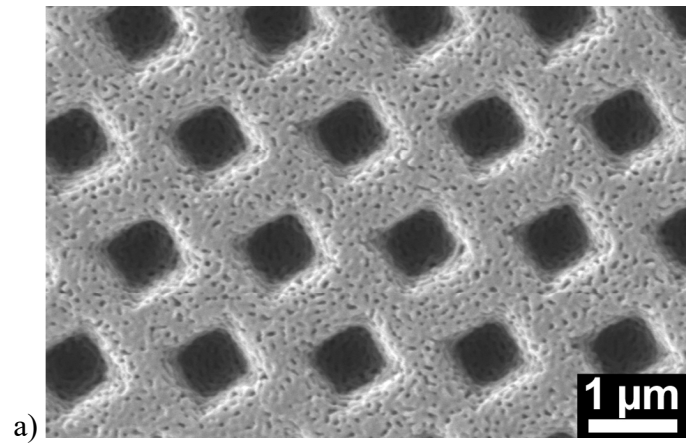

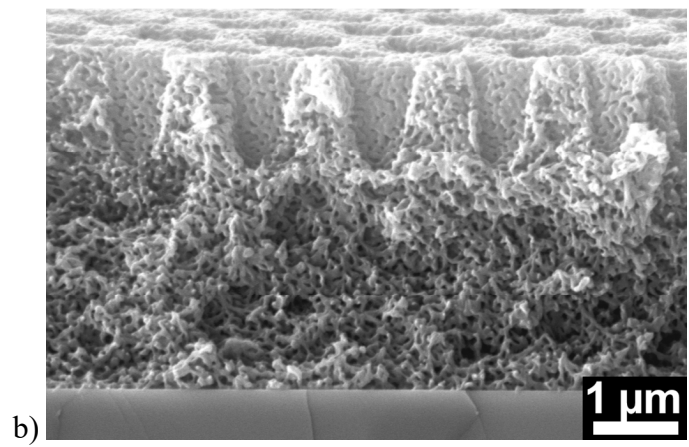

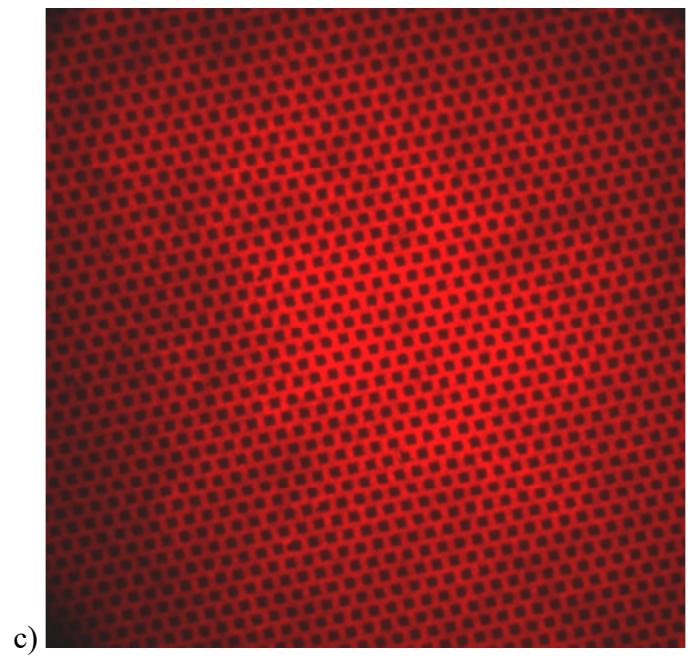

**Figure 4.** a), b) Scanning electron microscopy images of a spongy macroporous PS-*b*-P2VP monolith corresponding to panel i) of Figure 1. a) Top view; b) cross section showing the spongy macroporous PS-*b*-P2VP monolith on top and the underlying Si wafer at the bottom. c) TIRFM image of a glass slide patterned with a holey rhodamine B film by stamping an aqueous rhodamine B solution infiltrated into a spongy macroporous PS-*b*-P2VP monolith under ambient conditions. The displayed image shows the result of the 5$^{th}$ consecutive stamping cycle without reloading the spongy macroporous PS-*b*-P2VP monolith with ink. The displayed image field has an area of 53.91 µm × 53.91 µm.

A spongy macroporous PS-*b*-P2VP monolith with a size of 0.6 cm × 0.6 cm still connected to a Si wafer was used as stamp. The underside of the Si wafer was glued onto a stamp holder with a mass of 27 g in such a way that the macropores of the PS-*b*-P2VP monolith were exposed. The pressure imposed by the mass of the stamp holder was estimated to be 7.35 kN/m$^2$. In a first stamping experiment we used a solution of 0.05 wt-% of the dye rhodamine B in water as ink. We dropped 50 µL of the aqueous rhodamine B solution onto the macroporous surfaces of the PS-*b*-P2VP monoliths to fill the mesopore systems with ink (Figure 1j). After 1 min, residual ink was removed from the macroporous surface of the PS-*b*-P2VP monoliths with filter paper. Then, stamping on glass slides was instantly performed under ambient conditions with a contact time of 1 s per stamping cycle by bringing the macroporous surface of the spongy macroporous PS-*b*-P2VP monoliths into contact with the glass slides (Figure 1k). Supporting Figure S6 shows a total internal reflection fluorescence microscopy (TIRFM) image of the result of the first stamping cycle after inking, whereas Figure 4c displays the result of the 5$^{th}$ consecutive stamping cycle without re-inking. The patterns obtained in this way are faithful replicas of the macropore arrays of the spongy macroporous PS-*b*-P2VP monoliths (Figure 1l). The areas in which the walls surrounding the macropores contacted the underlying glass slides showed fluorescence, indicating that in these areas ink containing rhodamine B was transferred to the underlying glass slide. Conversely, at the positions of the macropores no rhodamine B was deposited onto the glass slides and no fluorescence can be seen. As a result, a continuous rhodamine B film was formed that contained a hexagonal array of quadratic holes with edge lengths of ~950 nm and areas of ~0.9 µm². Since the lattice constant of the hexagonal array amounts to 1.5 µm, the rhodamine B film covers 60 % of the surface of the glass slide. The comparison of Supporting Figure S6 and Figure 4c reveals that the quality of the stamped pattern did not deteriorate.

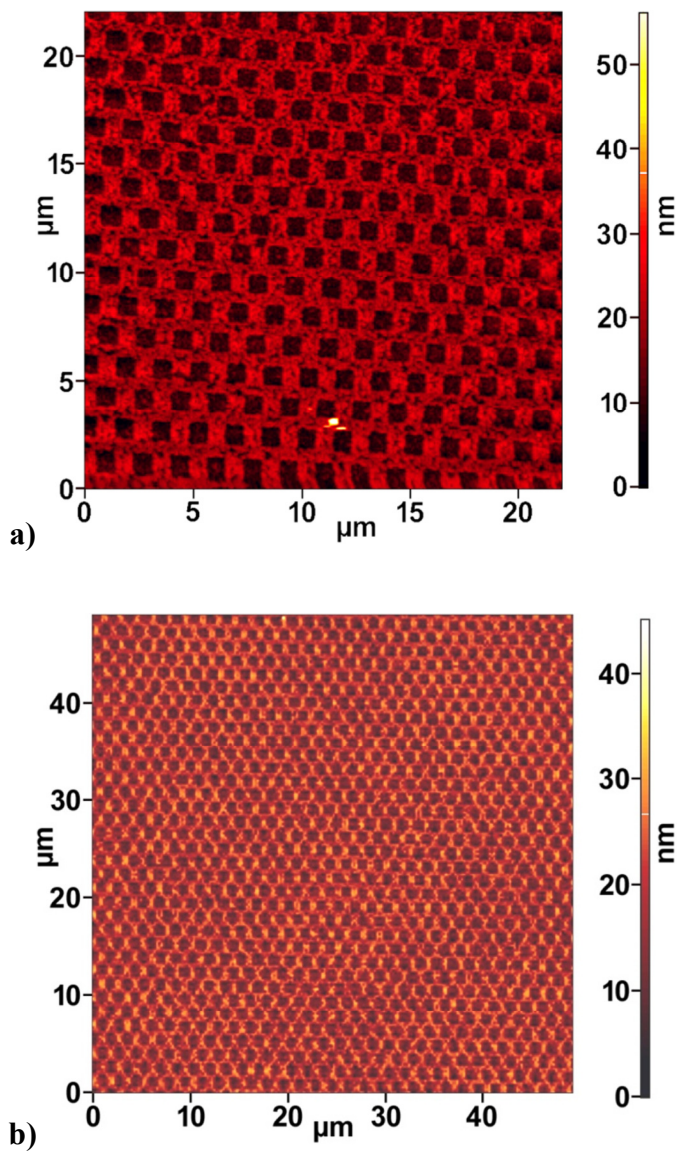

a)

b)

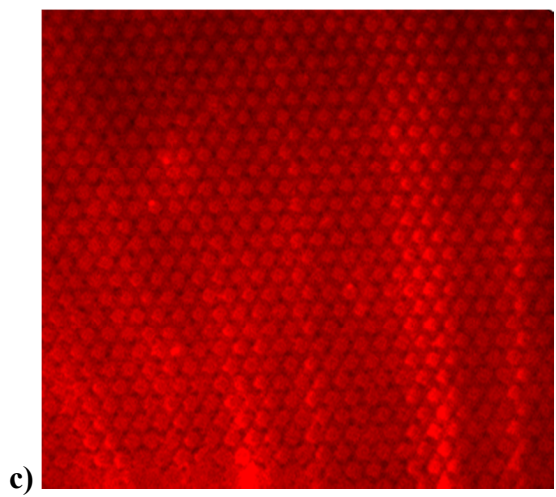

**Figure 5.** Stamping of 1-dodecanthiol films that contain holes at the positions of the macropores of the spongy macroporous PS-*b*-P2VP monoliths used as stamps onto gold-coated glass slides, followed by partial etching of the gold exposed in the holes not protected by 1-dodecanetiol. a) AFM topography image of a sample produced by the 1$^{st}$ stamping cycle; b) AFM topography image of a sample produced by the 5$^{th}$ consecutive stamping cycle. In panels a) and b) the dark dots represent the unprotected (partially etched) gold surface at the positions of the macropores of the spongy macroporous PS-*b*-P2VP monoliths, whereas 1-dodecanethiol was printed onto the continuous elevated areas separating the dark dots. c) TIRFM image of a sample prepared like that shown in panel a) after immersion into an aqueous solution of 0.05 wt-% rhodamine B for 30 s. The image field of panel c) has a width of 38.54 μm and a height of 36.12 μm.

In a second stamping experiment we stamped 1-dodecanethiol films with regular arrays of holes at the positions of the macropores of the spongy macroporous PS-*b*-P2VP monoliths used as stamps onto gold-coated glass slides. The spongy macroporous PS-*b*-P2VP monoliths were infiltrated with 20 μL of a 3 mM ethanolic 1-dodecanethiol solution After 30 s we removed residual solution from the macroporous surface of the spongy macroporous PS-*b*-P2VP monoliths with filter paper and started stamping with a contact time of 1 s per cycle immediately thereafter. The exposed gold not protected by adsorbed 1-dodecanethiol was partially etched following procedures reported elsewhere [18b, 25] to enhance contrast in the following microscopic investigations. As revealed by AFM measurements (Figure 5a,b), the thickness of the gold layer was reduced from initially 30 nm to 10-15 nm. As obvious from the AFM height images, the gold was etched in square-shaped dot-like areas with edge lengths of ~1.0 µm forming hexagonal arrays with a lattice constant of ~1.5 µm. Hence, no 1-dodecanethiol was deposited in the square-shaped dot-like areas that replicated the positions of the macropores of the spongy macroporous PS-*b*-P2VP monoliths. On the other hand, the continuous 1-dodecanethiol films surrounding the square-shaped dot-like areas covered ~56 % of the surface, which is consistent with the results obtained by stamping aqueous rhodamine B ink (Figure 4c). Immersion of thus-treated surfaces into an aqueous rhodamine B solution resulted in selective adsorption of the rhodamine B onto the exposed gold surfaces. Therefore, strong rhodamine B fluorescence occurred in the discrete dot-like areas not covered by 1-dodecanethiol at the positions of the macropores of the spongy macroporous PS-*b*-P2VP monoliths (Figure 5c).

**Conclusion**

We have explored non-elastomeric monoliths consisting of the block copolymer PS-*b*-P2VP having surfaces topographically patterned with regular arrays of micron-sized macropores as means to pattern substrates. The macroporous PS-*b*-P2VP monoliths were prepared by a double replication process. Ordered macroporous silicon containing wafer-scale macropore monodomains was used as primary mold. PDMS secondary molds with pillars as negative replicas of the macropores of the primary macroporous silicon molds were in turn molded against PS-*b*-P2VP solutions to obtain the macroporous PS-*b*-P2VP monoliths. The macroporous PS-*b*-P2VP monoliths were, therefore, positive replicas of the primary macroporous silicon molds and negative replicas of the secondary PDMS molds. Macroporous PS-*b*-P2VP monoliths enable modes of substrate patterning that go beyond the functionality of solid elastomeric stamps. In a first application example (also possible with macroporous monoliths consisting of the corresponding homopolymers) in each macropore of the PS-*b*-P2VP monoliths attached to underlying silicon wafers one NaCl nanocrystal was deposited. After coating the macroporous PS-*b*-P2VP monoliths with thin iridium layers the PS-*b*-P2VP was decomposed by heat treatment. The remaining iridium skins formed corrugated films on the silicon wafers that exhibited regular arrays of hexagonal cells replicating the macropores of the destroyed macroporous PS-*b*-P2VP monoliths. Exactly one NaCl crystal was located in each cell. In a second model application the macroporous PS-*b*-P2VP monoliths were further functionalized with a continuous spongy mespore system by selective-swelling induced pore generation. Using the spongy mesopore system as ink reservoir enabled capillary microstamping that yielded continuous ink films with holes at the positions of the macropores. Multiple capillary microstamping cycles could be carried out under ambient conditions without quality deterioration of the stamped patterns. These examples demonstrate that substrate patterning with non-elastomeric polymer monoliths – including the deposition of nanocrystals – may be employed to generate preconcentration sensors, devices for data storage and energy conversion, as well as surfaces with specific adhesion, wettability and tailored interactions with biological cells.

**Experimental Section**

*Materials:* Macroporous silicon with a pore diameter of 1 μm, a pore depth of 700 nm and a lattice constant of 1.5 μm was obtained from SmartMembranes (Germany). Cylinder-forming PS-*b*-P2VP ($M_n$(PS) = 101 kg/mol, $M_n$(P2VP) = 29 kg/mol, polydispersity = 1.6) was purchased from Polymer Source Inc. (Canada). Tetrahydrofuran (THF, ≥99.9%), ethanol (≥99.9%), rhodamine B (≥95%) and 1-dodecanethiol (≥98%) were obtained from Sigma Aldrich. PDMS prepolymer formulation (Sylgard® 184) was supplied by Dow Corning Corporation. $K_2S_2O_3$, KOH, $K_3Fe(CN)_6$ and $K_4Fe(CN)_6$ were obtained from Alfa Aesar; 1-octanol was purchased from Merck. All chemicals were used without further purification. Silicon wafers as the substrates to prepare the macroporous PS-*b*-P2VP monoliths were cut into the size of 2 cm × 2 cm, ultrasonicated at least three times in ethanol and dried before use.

*Preparation of macroporous PS-b-P2VP monoliths:* To prepare PDMS secondary molds, base and curing agent of the PDMS prepolymer formulation were mixed at a weight ratio of 10:1 under stirring. After degassing for 1 h under ambient conditions at room temperature, the PDMS prepolymer mixture was poured onto the macroporous silicon primary molds and cured at 70 ºC for 12 h. The PDMS secondary molds with a thickness of ~1 mm were detached from the macroporous silicon primary molds and cut into pieces extending 1 cm × 1 cm. A portion of 20 µL of a solution containing 5 wt-% PS-*b*-P2VP in THF was drop-cast on smooth silicon wafers at room temperature and instantly covered with the PDMS secondary molds under a pressure of 3.9 kN/m$^2$. After a wait time of 15 min the PDMS secondary molds were detached from the macroporous PS-*b*-P2VP monoliths obtained in this way.

*Use of macroporous PS-b-P2VP monoliths as sacrificial templates:* The macroporous PS-*b*-P2VP monoliths were immersed into aqueous solutions containing 2 mol/L NaCl for 1 min followed by drying for 2 h under ambient conditions at room temperature. The macroporous PS-*b*-P2VP monoliths containing one NaCl nanocrystal per macropore were sputter-coated with a ~5 nm thick iridium layer (Sputter Angle Emitech K575X, UK). The thickness of the iridium layer was determined by AFM (see below). Thermal decomposition of the PS-*b*-P2VP was carried out by heating the samples to 540°C at a rate of 1 K/min followed by annealing at this temperature for 3 h in air.

*Capillary microstamping:* To fabricate spongy macroporous PS-*b*-P2VP monoliths, solid macroporous PS-*b*-P2VP monoliths still attached to underlying Si wafers were immersed into ethanol at 70 ºC for 30 min followed by drying under ambient conditions. The spongy macroporous BCP monoliths along with the underlying Si wafers were cut into pieces extending 0.6 cm × 0.6 cm and glued onto cylindrical sample holders with a weight of 27 g in such a way that the macropores of the spongy macroporous PS-*b*-P2VP monoliths were exposed. To stamp rhodamine B, 50 µL portions of aqueous solutions containing 0.05 wt-% rhodamine B were drop-cast onto the spongy macroporous PS-*b*-P2VP monoliths within 1 min. Residual solution was removed using filter paper. Then, the spongy macroporous PS-*b*-P2VP monoliths were instantly brought into contact with glass slides for ~1 s. To stamp 1-dodecanethiol, glass slides were coated with a ~7 nm thick titanium layer and a ~30 nm thick gold layer using a Balzers BAE 120 evaporator applying recipes reported elsewhere.[18b] 20 µL portions of a 3 mM 1-dodecanethiol solution in ethanol were infiltrated into the spongy macroporous PS-*b*-P2VP stamps. Residual solution was removed using filter paper after 30 s. Stamping onto the gold-coated glass slides was performed immediately thereafter with a contact time of ~1 s. To partially etch the gold layer, 0.248 g $K_2S_2O_3$, 0.261 g KOH, 0.032 g $K_3Fe(CN)_6$, 0.0042 g $K_4Fe(CN)_6$ and 3.18 µL 1-octanol were mixed with 10 mL deionized water. Gold-coated glass slides onto which holey 1-dodecanethiol films had been stamped were immersed into the thus-obtained solution for 6 min followed by rinsing with deionized water and ethanol. To adsorb rhodamine B onto samples treated in this way, the samples were immersed into an aqueous solution of 0.05 wt-% rhodamine B for 30 s.

*Characterization:* Scanning electron microscopy was carried out using a Zeiss Auriga device operated at an accelerating voltage of 7 kV. Before SEM investigations the samples were sputter-coated with a ~5 nm thick iridium layer. AFM was carried out using a NTEGRA microscope (NT-MDT). TIRF-Microscopy was carried out with an inverted microscope (Olympus IX71) equipped with a 4-Line TIRF condenser (Olympus TIRF 4-Line). A 150-fold oil immersion Objective (Olympus UAPON 150x TIRF, NA 1.45) was used in combination with a 561 nm diode-pumped solid state laser (Cobolt Jive 561). A TIRF pentaband polychroic beamsplitter (Semrock zt405/488/561/640/730rpc) and a pentabandpass emitter (BrightLine HC 440/521/607/694/809) were used. Images were acquired by an electron multiplying CCD camera (Andor iXon Ultra 897) and processed using Olympus CellSens Software.

**Supporting Information**
Supporting Information is available from the author.

**Acknowledgements**
The authors thank the European Research Council (ERC-CoG-2014, project 646742 INCANA) for funding.